# Bifurcations of homoclinic orbits in bimodal maps

Kai T. Hansen
*Dep. of Physics, University of Oslo, Box 1048, Blindern, N-0316 Oslo, Norway*

We discuss the bifurcation structure of homoclinic orbits in bimodal one dimensional maps. The universal structure of these bifurcations with singular bifurcation points and the web of bifurcation lines through the parameter space are described. The bifurcations depend on two parameters (codimension 2 bifurcations). We find the bifurcation lines exactly in a symbolic dynamics parameter plane and numerically in the parameter planes of a polynomial map and a piecewise linear map.

The aim of this article is to describe the structure of the bifurcations of some non-periodic orbits in a bimodal map. A more complete discussion of the bifurcations of orbits in bimodal, trimodal and four-modal maps will be given in ref. [1] (see also Ref. [2]). Bifurcations of periodic orbits in bimodal maps are investigated by several authors [3–7].

The bimodal map is a one dimensional continuous map $x_{t+1} = f(x_t; a, b)$ where the function $f(x)$ have one maximum point $f(x_{c_1})$, and one minimum point $f(x_{c_2})$, and two parameters $(a, b)$ such that the two extremum points can be changed independently. We call $x_{c_1}$ and $x_{c_2}$ the critical points of the map. If $x_{c_1} < x_{c_2}$ we denote the map $+-+$ and if $x_{c_1} > x_{c_2}$ we denote it $-+-$. In this letter we choose only to study the map $+-+$ while the $-+-$ map will be described elsewhere [1].

Periodic orbits are solutions of the equation

$$g(x) = f^{(n)}(x) - x = 0.$$

The solution of $g(x) = 0$ is the topic of singularity theory [8] and the bifurcation of periodic orbits in the bimodal maps have codimension 2 structure, i.e. a universal form in a two-parameter plane. Non-periodic orbits are solutions of a slightly different equation and yields a different bifurcation picture but also with codimension 2.

We choose the polynomial map

$$x_{t+1} = x_t^3 - ax_t + b \tag{1}$$

as an example of a smooth bimodal map, and as a continuous piecewise linear map we choose

$$x_{t+1} = \begin{cases} 2x_t + a + b + 2 & \text{if } x_t \le -1 \\ b - ax_t & \text{if } -1 < x_t \le 1 \\ 2x_t - a + b - 2 & \text{if } x_t > 1 \end{cases} \tag{2}$$

These two very different bimodal maps illustrate which structures that are common for all bimodal maps, and the changes of the structure when the smoothness of $f(x)$ is different.

We define symbolic dynamics [2,7,9–12] for the bimodal map $s_t \in \{0, 1, 2\}$

$$s_t = \begin{cases} 0 & \text{if } x_t \le x_{c_1} \\ 1 & \text{if } x_{c_1} < x_t \le x_{c_2} \\ 2 & \text{if } x_t > x_{c_2} \end{cases} \tag{3}$$

An orbit is identified by an infinite symbol string $S = \ldots s_{-2}s_{-1}s_0s_1s_2\ldots$ and the points $\ldots, x_{-2}, x_{-1}, x_0, x_1, x_2, \ldots$ in an unstable orbit are uniquely defined by the symbol string. A point $x_t$ of an orbit $S$ has one symbolic future $s_{t+1}s_{t+2}s_{t+3}\ldots$ and one symbolic past $\ldots s_{t-2}s_{t-1}s_t$. Since the bimodal map is not uniquely invertible a point $x_t$ can belong to different orbits and have different symbolic pasts, but the point always has one unique symbolic future. We use the convention $\overline{s_1s_2\ldots s_n} = (s_1s_2\ldots s_n)^\infty$ to denote an infinite repetition of a finite length symbol string. We also use the notation $\ldots s_{n-1}\{s_n, s'_n\}s_{n+1}\ldots$ for the two symbol strings $\ldots s_{n-1}s_ns_{n+1}\ldots$ and $\ldots s_{n-1}s'_ns_{n+1}\ldots$.

It is useful to define the following variable for the $+-+$ bimodal map, following ideas of Milnor and Thurston [2,11,13].

$$
\begin{aligned}
w_1 &= s_1 \\
p_1 &= \begin{cases} 1 & \text{if } s_1 = 0 \text{ or } 2 \\ -1 & \text{if } s_1 = 1 \end{cases} \\
w_t &= \begin{cases} s_t & \text{if } p_{t-1} = 1 \\ (2-s_t) & \text{if } p_{t-1} = -1 \end{cases} \\
p_t &= \begin{cases} p_{t-1} & \text{if } s_t = 0 \text{ or } 2 \\ -p_{t-1} & \text{if } s_t = 1 \end{cases} \\
\tau(x_0) &= 0.w_1 w_2 w_3 \ldots = \sum_{t=1}^{\infty} \frac{w_t}{3^t}
\end{aligned}
\tag{4}
$$

where the base 3 real number $0 \leq \tau \leq 1$ is called the symbolic value. The kneading values are defined as

$$
\begin{aligned}
\kappa_1 &= \tau(x_{c_1}) \\
\kappa_2 &= \tau(x_{c_2})
\end{aligned}
\tag{5}
$$

Let $x_t$, $t \in \{\ldots, -2, -1, 0, 1, 2, \ldots\}$ be the points of an orbit specified by the symbol string $S$ and let the maximum and minimum values of the symbolic value be given by

$$
\begin{aligned}
\tau_1^{\max}(S) &= \max_t \tau(x_t) \quad \text{with} \quad x_t < x_{c_2} \\
\tau_2^{\min}(S) &= \min_t \tau(x_t) \quad \text{with} \quad x_t > x_{c_1}
\end{aligned}
\tag{6}
$$

The orbit $S$ is admissible if and only if

$$
\begin{aligned}
\kappa_1 &> \tau_1^{\max} \\
\kappa_2 &< \tau_2^{\min}
\end{aligned}
\tag{7}
$$

The values $\kappa_1$ and $\kappa_2$ depend on the parameters $a$ and $b$ and we denote them the symbolic parameters of the bimodal map and the plane $(\kappa_1, \kappa_2)$ the symbolic parameter plane. An orbit $S$ is admissible in a rectangle in the $(\kappa_1, \kappa_2)$ parameter plane. At the symbolic parameter values $\kappa_1 = 1.0$ and $\kappa_2 = 0.0$ all orbits in the map exists. With these tools we can analyze the bifurcations of all orbits in the bimodal map.

For simplicity we will here discuss only one family of homoclinic orbits. Other non-periodic orbits will have similar structures and these orbits can be analyzed in the same way.

A homoclinic orbit in a bimodal map has a symbolic description

$$S_{A,B} = \overline{A}B\overline{A}$$

where $A$ and $B$ are finite symbol strings. We choose the simplest case with $A = 1$

$$S_{1,B} = \overline{1}B\overline{1} \tag{8}$$

with $B = b_1 b_2 \ldots b_n$.

The bifurcation lines of $S_{1,B}$ with a few different strings $B$ are drawn in the symbolic parameter plane $(\kappa_1, \kappa_2)$ in Fig. 1. The two homoclinic orbits which exist for the largest area in the parameter plane are $\overline{1}2\overline{1}$ and $\overline{1}212\overline{1}$ (in the notation introduced above $\overline{1}\{1,2\}12\overline{1}$) which both have $\tau_1^{\max} = \tau(2\overline{1}) = 0.2\overline{1} = 5/6$ and $\tau_2^{\min} = \tau(12\overline{1}) = 0.10\overline{1} = 7/18$ yielding the bifurcation lines $\kappa_2 = 0.2\overline{1}$ and $\kappa_2 = 0.10\overline{1}$.

The homoclinic orbit is drawn in Fig. 2 (a) with $a = 2.00794$, $b = 0.274$ that are parameters close to the singular point in the parameter plane where both the critical points $x_{c_1}$ and $x_{c_2}$ belongs to the homoclinic orbit. In contrast to the bifurcation of periodic orbits is it important to notice which critical point that is first and which that is last along the orbit at the singular parameter point. In Fig. 2 (a) the orbit is first close to $x_{c_2}$ then close to $x_{c_1}$. There are four orbits bifurcating at this singular point with the symbolic description

$$\overline{1}\{1,2\}\{0,1\}2\overline{1} \tag{9}$$

The structure of the bifurcation lines from the singularity depends on whether the line is associated with the first or the last critical point. The last critical point that the orbit visits yields one bifurcation line for all four orbits. In this example the line is $\kappa_1 = 0.2\overline{1}$ in the symbolic parameter space. The critical point first visited by the orbit yields two different bifurcation lines; $\kappa_2 = 0.10\overline{1}$ for the orbits $\overline{1}\{1,2\}12\overline{1}$ and $\kappa_2 = 0.02\overline{1} = 5/18$ for the two orbits $\overline{1}\{1,2\}02\overline{1}$. These three lines are drawn in the symbolic parameter plane in Fig. 1 (a). The two $\kappa_2$ lines hit the $\kappa_1$ line at two different points, but these two points can be identified with one singular point because the line segment $\kappa_1 = 0.2\overline{1}$, $0.02\overline{1} < \kappa_2 < 0.10\overline{1}$ is not accessible.

The bifurcation lines in Fig. 1 (a) are drawn in the $(a, b)$ plane in Fig. 3 (a) for the polynomial map (1). The main difference from the symbolic parameter plane is that the two bifurcation lines corresponding to the symbolic parameter lines $\kappa_2 = 0.02\overline{1}$ and $\kappa_2 = 0.10\overline{1}$ hit the line corresponding to $\kappa_1 = 0.2\overline{1}$ approximately like the two branches of a parabola hit tangentially the end-points of a straight line. The area in the parameter space which gives a specific solution close to a singularity is often called a cusp because of the shape. In this case we have two different cusps; one which is a narrow sharp cusp and one which does not look like a cusp at all. The boundary lines of the second cusp have a common first derivative at the singularity but higher derivatives differ; the curve is $C^1$ but not $C^2$. We believe that this is generic for smooth bimodal maps with $f''(x_{c_i}) \neq 0$. A simple way of understanding the cusp structures is to look at the solution of the simple equation

$$f_b(f_a(x)) = 0 \tag{10}$$

where $f_a(x)$ and $f_b(x)$ are functions with one extremum value each. Choosing parabolas $f_a(x) = x^2 - a$ and $f_b(x) = x^2 - b$ gives

$$(x^2 - a)^2 - b = 0$$

which has four real solutions for $a > 0$ and $0 < b < a^2$ (a sharp cusp) and which yields at least two real solutions in the area bounded by $b > 0$ when $a \geq 0$ and $b > a^2$ when $a < 0$ (a smooth cusp).

The bifurcation lines for the piecewise linear map are drawn in Fig. 4 (a). Here the three bifurcation lines simply hit the singular point with different angles giving two cusps where the derivatives of the border lines are different at the singularity. The bifurcation structure of the piecewise linear map is related to the solution of (10) when the functions $f_a$ and $f_b$ are piecewise linear. Choosing $f_a(x) = |x| - a$ and $f_b(x) = |x| - b$ gives

$$||x| - a| - b = 0$$

and this equation has four real solutions for $0 < b < a$ and it has at least two real solutions in the area bounded by $b > 0$ when $a \geq 0$ and $b > -a$ when $a < 0$. These are two cusps of the same type as the cusps in Fig. 4 (a).

Equation (10) captures the essence of a bifurcation where an orbit is born when this orbit does not have any point close to the critical points in an arbitrarily long future after visiting the two critical points once. The function $f_b(f_a(x))$ yields the same structure as obtained locally around an orbit visiting in any order: two times close to independent extremum points, and a finite number of times at points where the function is monotonically increasing or decreasing. An orbit is not close to any critical point (e.g. an unstable periodic orbit) can be considered to have constant points close to the singularity, and we choose a constant point in this orbit to be zero. The bifurcation of periodic orbits is different and is related to the solutions of an equation like

$$f_b(f_a(x)) = x$$

which gives different structures. The borderlines of a typical cusp for periodic orbits separates proportional to the distance from the cusp to the power 3/2 for smooth maps [8] compared with the narrower cusp of equation (10) with smooth functions where the distance between the borderlines increases as the distance from the singularity to the power 2.

So far we have found that the structure of the singular point is different for periodic orbits and for homoclinic orbits. A further difference is that two singular bifurcation points for periodic orbits are always separated by a finite distance both in the symbolic parameter plane and in the parameter plane of a smooth map, while singular bifurcation points of homoclinic orbits may be arbitrarily close and may have bifurcation lines in common.

In Fig. 1 (b) the bifurcation lines for the four homoclinic orbits

$$\overline{1}\{0,1\}2\{1,2\}02\overline{1} \tag{11}$$

are added in the symbolic parameter plane. These four orbits have the bifurcation line $\kappa_2 = 0.02\overline{1}$ in common with the orbits $\overline{1}\{1,2\}02\overline{1}$ but have different $\kappa_1$ bifurcation lines; $\kappa_1 = 0.2120\overline{1} = 139/162$ for the two orbits $\overline{1}\{0,1\}2102\overline{1}$ and $\kappa_1 = 0.2202\overline{1} = 149/162$ for the two orbits $\overline{1}\{0,1\}2202\overline{1}$. This yields a singular bifurcation point with three bifurcation lines but with $\kappa_1$ and $\kappa_2$ switched compared to the structure of the bifurcation discussed above, because at this singular point the orbit (11) visits the point $x_{c_1}$ before it visits $x_{c_2}$. The homoclinic orbit close to the singular parameter value of the polynomial map is drawn in Fig. 2 (b) with $a = 2.34865$ and $b = 0.26552$.

The structure of the bifurcation lines are drawn in the $(a, b)$ plane of the polynomial map (1) in Fig. 3 (b) and the structure of the singularity is the same as for the orbits (9). The bifurcation lines for the piecewise linear map are drawn in Fig. 4 (b). This latter singular point is a finite distance away from the singularity discussed above but choosing for example the homoclinic orbits $\overline{1}\{0,1\}21^{(2n)}\{1,2\}02\overline{1}$ these will yield singular bifurcation points converging to the singular bifurcation point of the orbit $\overline{1}\{1,2\}\{0,1\}2\overline{1}$ since these have $\kappa_1 = 0.21^{(2n+1)}20\overline{1}$ or $\kappa_1 = 0.21^{(2n)}202\overline{1}$ which in the limit $n \rightarrow \infty$ is $\kappa_1 = 0.2\overline{1}$. These singular points will be arbitrarily close also in the $(a, b)$ plane of the polynomial map and the piecewise linear map. The set of all singular bifurcation points on the line $\kappa_2 = 0.02\overline{1}$ is a complicated set of the Cantor type. Since each new bifurcation line in the Cantor set of lines again has new singular points in a Cantor set etc. this yields a complicated web of crossing bifurcation lines through the parameter plane connecting the homoclinic orbits of the form $\overline{1}B\overline{1}$.

If the homo-/hetroclinic orbit ends in a periodic orbit which is created in a tangent bifurcation, for example a period 3 orbit, then the cusp structure described above take place at the tangent bifurcation line in a tail of the swallowtail of the stable orbit.

The parameter values when the critical point maps into an unstable orbit in the unimodal map is called Misiurewicz point and corresponds to the bifurcation lines described here. Some of these bifurcations corresponds to major changes in the structure of the attractor, such as band merging bifurcations and crisis bifurcations. To understand how the chaotic attractor changes with the parameters these bifurcation lines has to be described.

In summary we have studied one example of the bifurcation structure for homoclinic orbits in bimodal maps. We find singular bifurcation points with cusps connected in a web of bifurcation lines through the parameter space. This is generic for all bimodal maps while the exact shape of the cusp depends on the smoothness of the function at the critical points. The bifurcation lines is found in a simple and exact way in the symbolic parameter space. The bifurcation structure for the homoclinic bifurcations in the two-dimensional Hénon map is similar but not identical to what is obtained here and this is discussed in [14]

The author is grateful for financial support from the Norwegian Research Council (NFR), and for interesting discussions with Predrag Cvitanović, Jan Frøyland and Tore M. Jonassen.

---

FIG. 1. Bifurcation lines for homoclinic orbits in the symbolic parameter plane.

FIG. 2. Homoclinic orbits in the polynomial map (1) close to a singular bifurcation point, (a) $\overline{1}\{1,2\}\{0,1\}2\overline{1}$ for $a = 2.00794$ $b = 0.274$, (b) $\overline{1}\{0,1\}2\{1,2\}02\overline{1}$ for $a = 2.34865$ $b = 0.26552$.

FIG. 3. Bifurcation lines in the parameter plane for the polynomial map (1).

FIG. 4. Bifurcation lines in the parameter plane for the piecewise linear map (2).

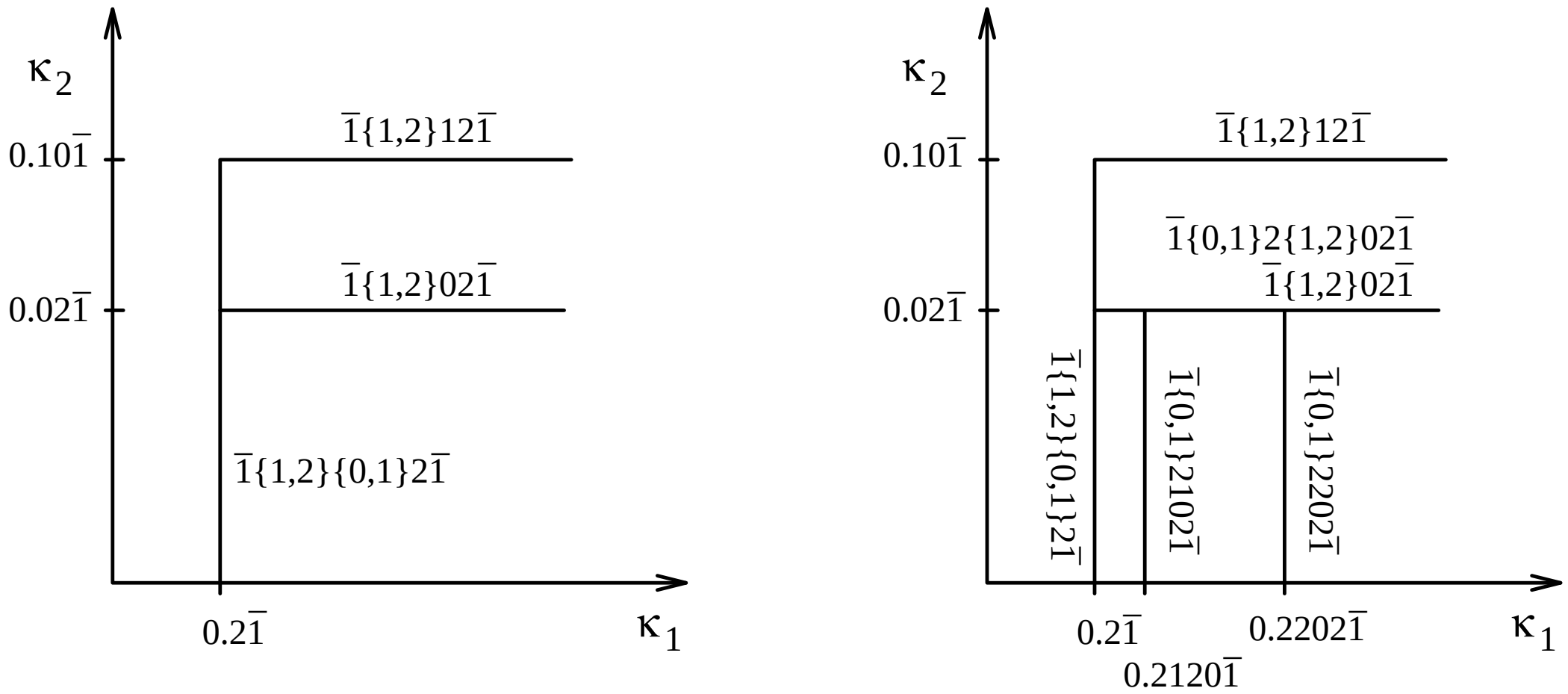

a) b)

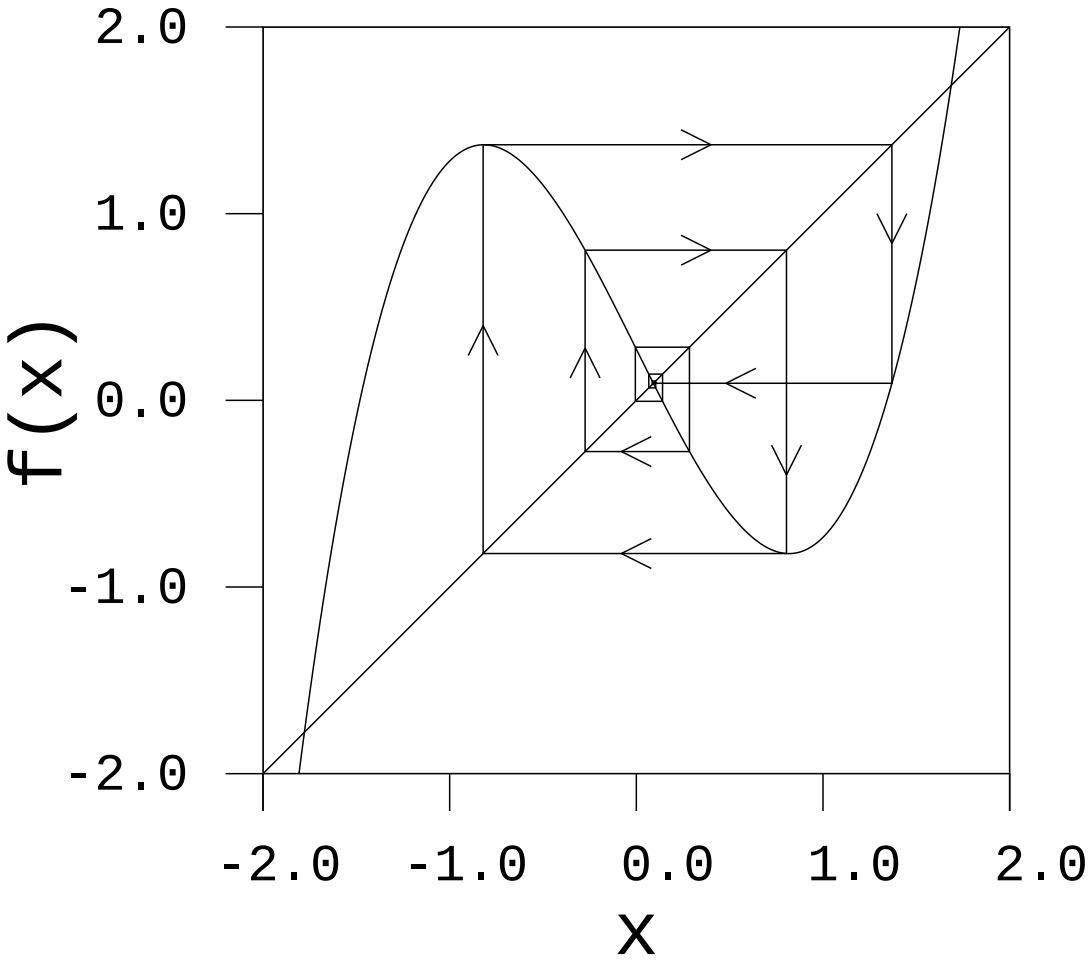

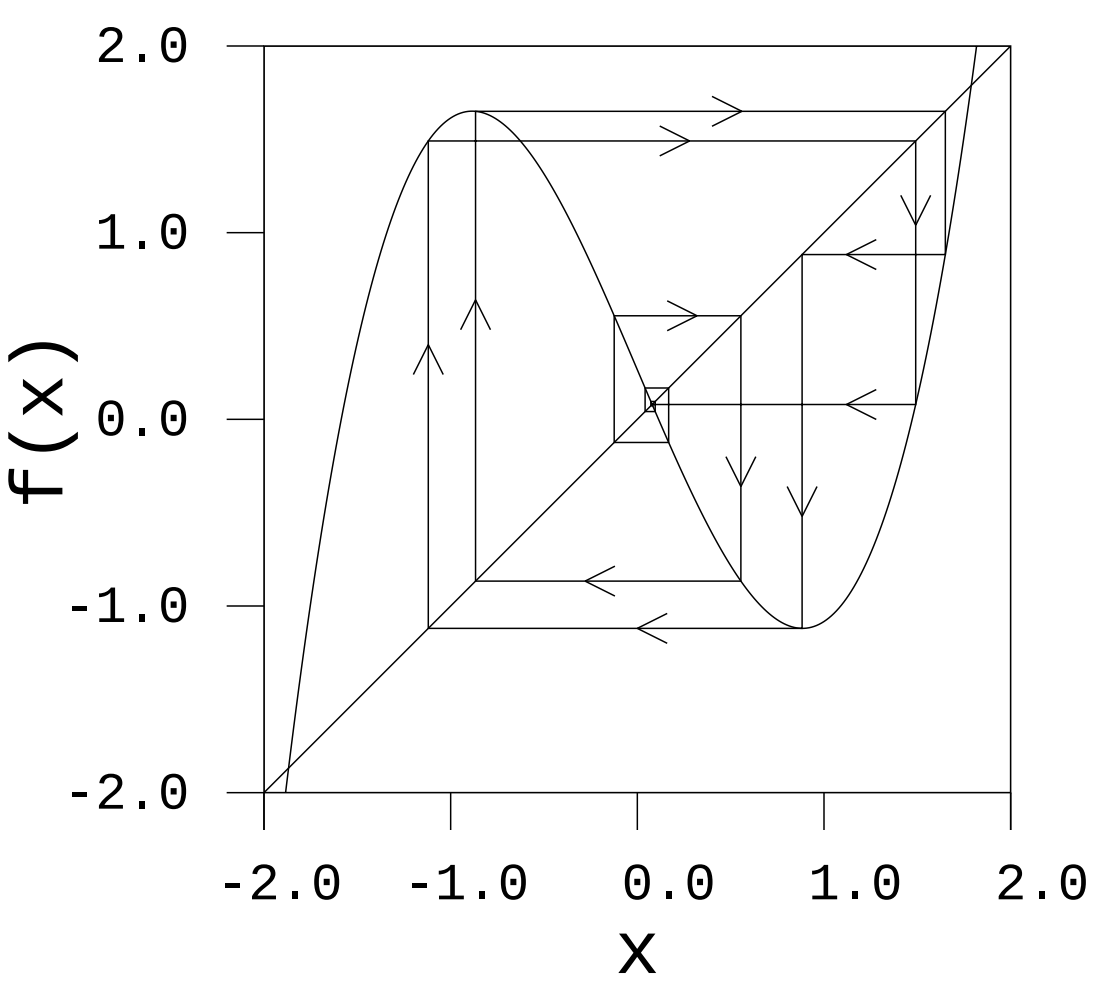

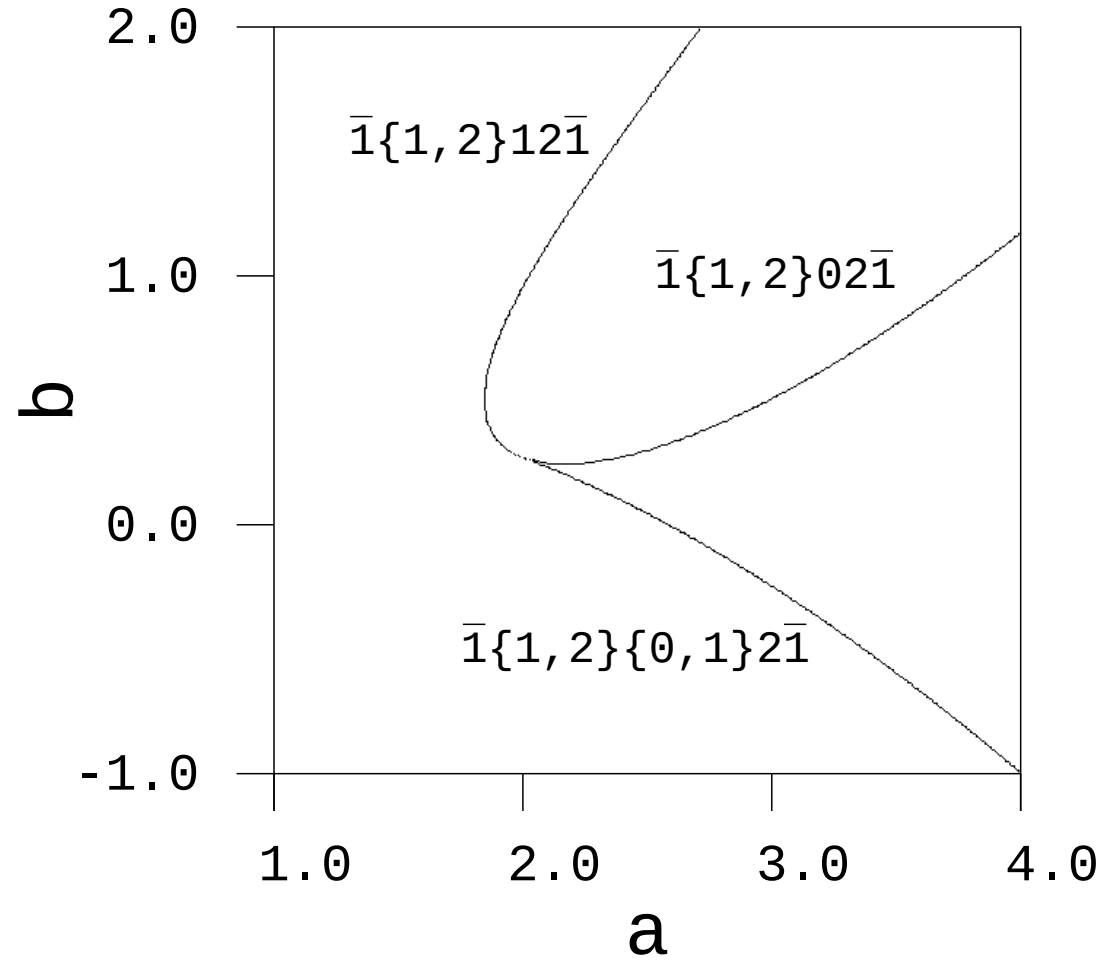

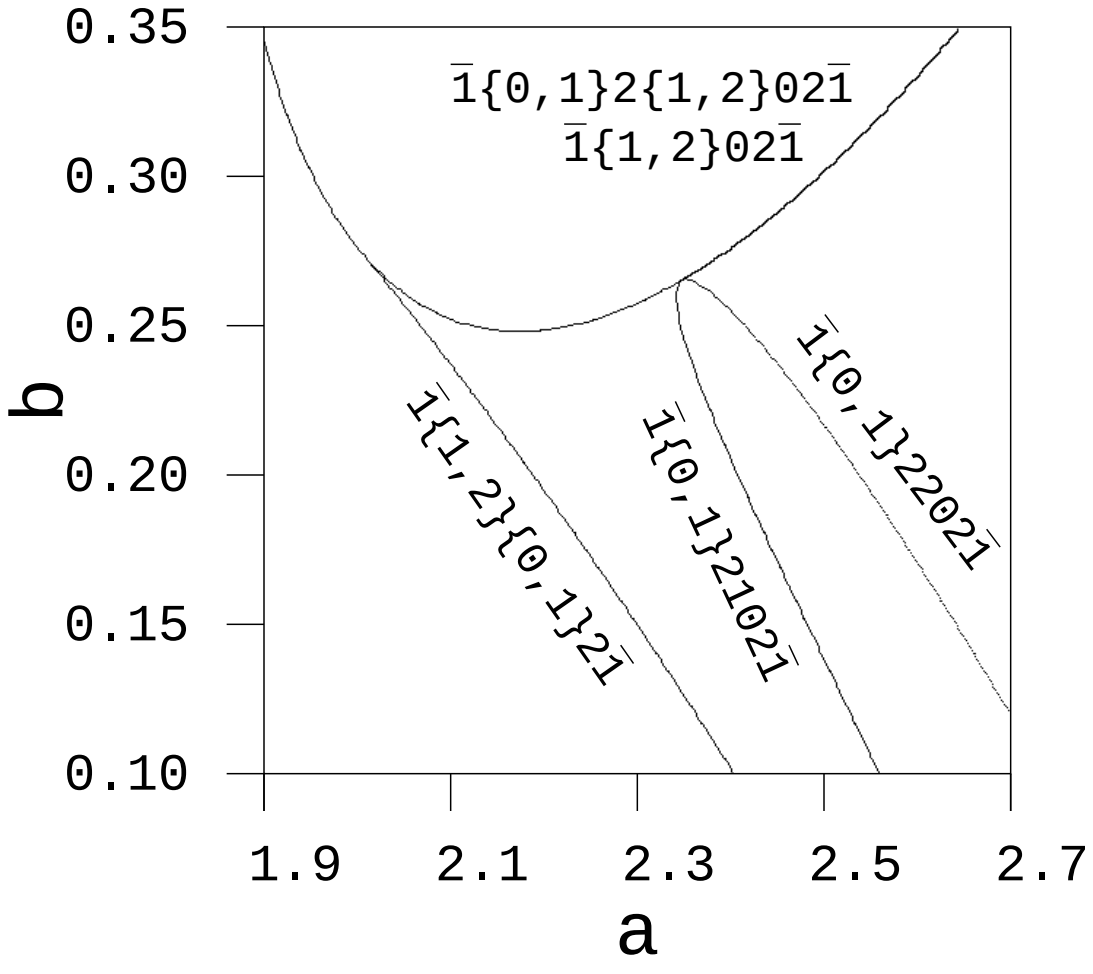

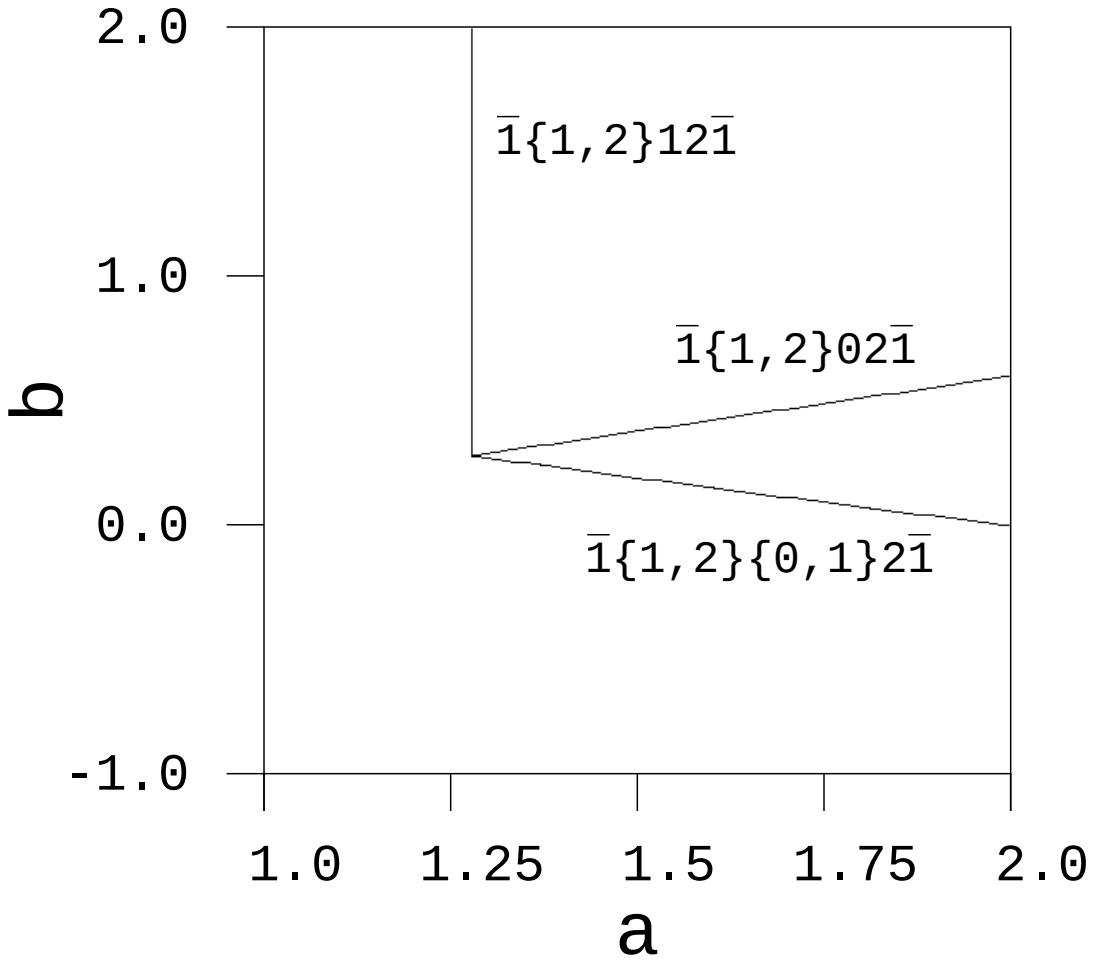

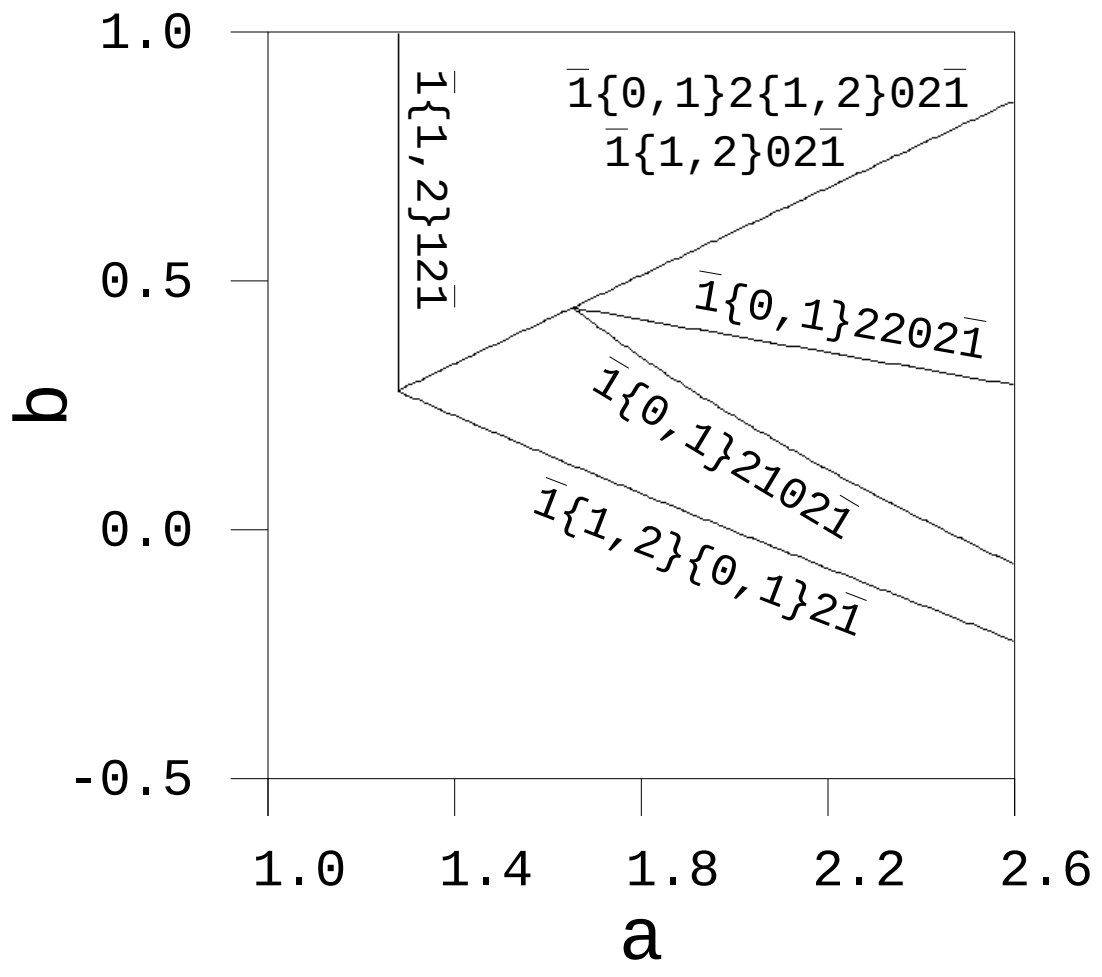